\pdfoutput=1
\documentclass{iopart} 
\usepackage{amssymb}
\usepackage{iopams} 
\usepackage{upgreek}
\usepackage{bm}
\usepackage{graphicx}
\eqnobysec
\begin{document}
\title{Electromagnetic self-force on a static charge in
  Schwarzschild-de Sitter spacetimes}      
\author{Joseph Kuchar$^1$, Eric Poisson$^1$, and Ian Vega$^{1,2}$} 
\address{$^1$ Department of Physics, University of Guelph, Guelph, Ontario,
N1G 2W1, Canada}
\address{$^2$ SISSA, Via Bonomea 265, 34136, Trieste, Italy}  
\ead{jkuchar@uoguelph.ca}
\ead{epoisson@uoguelph.ca} 
\ead{ian.vega@sissa.it}
\date{July 25, 2013} 

\begin{abstract} 
We compute the self-force acting on an electric charge at rest in 
Schwarzschild-de Sitter spacetimes, allowing the cosmological
constant to be either positive or negative. In the case of a positive
cosmological constant, we show that the self-force is always positive,
representing a repulsion from the black hole, and monotonically
decreasing with increasing distance from the black hole. The spectrum
of results is richer in the case of a negative cosmological
constant. Here the self-force is not always positive --- it is
negative when the black-hole and cosmological scales are comparable
and the charge is close to the black hole --- and not always monotonically
decreasing --- it is actually monotonically increasing when the
cosmological scale is sufficiently small compared to the black-hole
scale. The self-force also approaches a constant asymptotic value when
the charge is moved to large cosmological distances; this feature can
be explained in terms of an interaction between the charge and the
conformal boundary at infinity, which acts as a grounded conductor. 
\end{abstract} 

\section{Introduction} 
\label{sec:intro} 

To hold a particle in place in the static spacetime of a massive body 
requires an external force, which compensates for the body's
gravity. For a given spacetime, the required force depends on the
nature of the particle; in particular, the force differs when the
particle carries an electric charge, relative to its value when the
particle is neutral. The difference accounts for the particle's 
self-force, an effect that originates in the way that the particle's
electromagnetic field interacts with the spacetime curvature. This
interaction is subtle, it involves remote portions of the spacetime,  
and as a result the self-force is notoriously unintuitive; it is
virtually impossible to predict even the sign of the self-force before
it is revealed by a detailed calculation. For example, it is known
from the pioneering work of Smith and Will \cite{smith-will:80} that
the self-force on a particle at rest in the Schwarzschild spacetime of
a nonrotating black hole is repulsive, which implies a smaller
external force relative to a neutral particle; this outcome has defied
attempts to provide an intuition. (For a valiant attempt, refer to
Sec.~IV of Ref.~\cite{burko-etal:00}.)  

The study of self-forces in curved spacetime was initiated by DeWitt
and Brehme \cite{dewitt-brehme:60}, and there is currently a large
effort (reviewed in Ref.~\cite{poisson-pound-vega:11}) devoted to the
computation of self-forces in various spacetimes. Most of the recent
activity has focused on the gravitational self-force, in an effort to
model the inspiral and gravitational-wave emissions of extreme
mass-ratio binaries \cite{barack-sago:10, diener-etal:12,
  warburton-etal:12}. Our concern here is elsewhere. We have
unfinished business with even the simplest exemplars of self-forces in 
curved spacetimes, those involving static particles outside massive
bodies. We are troubled by the lack of intuition surrounding
these simple situations, and have pursued a line of inquiry that
aims to probe more deeply into the mysterious aspects of the
self-force. We intend to achieve this by producing a larger catalogue
of self-forces, examining its entries carefully, and extracting
whatever intuition can be extracted.    

Previous studies in this program have focused on altering the nature
of the massive body, replacing the black hole of the Smith-Will work
with a material body of some kind. Thus, Burko, Liu and Soen
\cite{burko-etal:00}, building on earlier work by Unruh
\cite{unruh:76}, replaced the black hole by a massive thin
shell. Shankar and Whiting \cite{shankar-whiting:07} considered a body
of constant density, and Isoyama and Poisson \cite{isoyama-poisson:12}
calculated the self-force outside relativistic polytropes. In all
cases the results confirmed the analysis of Drivas and Gralla
\cite{drivas-gralla:11}, who considered static charges in the exterior 
spacetime of a spherical body of arbitrary composition. In particular,
Drivas and Gralla expanded the self-force in powers of $r_0^{-1}$,
where $r_0$ is the radial position of the charge, and showed that the
leading term at order $r_0^{-3}$ is universal, independent of the
body's internal composition. The universal self-force is given by the
Smith-Will expression $e^2 M/r_0^3$, where $e$ is the particle's
charge and $M$ the mass of the body. They also showed that the
dependence of the self-force on details of internal composition
is revealed by the subdominant terms at order $r_0^{-5}$ and beyond;
these were examined closely by Isoyama and Poisson, who showed how the
self-force can be used as a probe of the body's internal structure.    

In this paper we place our focus elsewhere. Instead of altering the
nature of the massive body, we alter the nature of the asymptotic
spacetime. We calculate the self-force acting on a charged particle
held in place outside a black hole in a de Sitter universe. We
allow the cosmological constant to be either positive (de Sitter
asymptotics) or negative (anti de Sitter asymptotics). So while the
massive body is now fixed to be a black hole, the spacetime is no
longer asymptotically flat, and our goal is to determine the impact of
this change on the electromagnetic self-force. 

The immersion of the black hole in a de Sitter universe produces a
spacetime with two fundamental length scales. The first is the
event-horizon radius $r_e$, and the second is the cosmological scale
$r_c$; in the case of the Schwarzschild-de Sitter spacetime, $r_c$ is
also the radius of the cosmological horizon. The ability to tune
$r_c$ in relation to $r_e$ offers interesting possibilities for
self-force calculations, and broadens the range of situations for
which one might seek an intuition. To begin we may set $r_e \ll r_c$
and consider spacetimes for which there is a clean separation of
scales. In such a situation we might expect that the self-force is
mostly sensitive to the Schwarzschild aspects of the spacetime when
$r_0 \ll r_c$, and mostly sensitive to its de Sitter aspects when 
$r_0 \gg r_e$. 

In the case of Schwarzschild-de Sitter spacetime, this means that the
self-force should reduce to the Smith-Will expression when 
$r_0 \ll r_c$, while it should approach zero when $r_0 \gg r_e$,
possibly faster than $r_0^{-3}$. This is because de Sitter spacetime
is conformally flat and the electromagnetic field is conformally
invariant, leading to a vanishing self-force in pure de Sitter
spacetime. The situation is more subtle in Schwarzschild-anti de
Sitter spacetime, because of the presence of a conformal boundary at   
$r=\infty$. When $r_0 \ll r_c$ the self-force should still be well
approximated by the Smith-Will expression, but when $r_0 \gg r_e$ it 
should be conformally related to the force experienced by a particle
in a truncated flat spacetime.    

It is good news for intuition-building that these expectations are
borne out by our computations. The interesting cases, however,
are produced when the scales $r_e$ and $r_c$ are not widely
separated. In such a situation there is no Schwarzschild regime 
when $r_0 \ll r_c$, and there is no de Sitter regime when 
$r_0 \gg r_e$. Instead the self-force can be expected to depend on all
aspects of the spacetime, and one would be hard-pressed to predict the
outcome of a detailed computation. And indeed, we shall find some
surprising features when $r_e$, $r_0$, and $r_c$ are all comparable to
each other.

We explain how the electromagnetic self-force is computed in
Sec.~\ref{sec:methods}, and describe the results of our numerical   
calculations in Sec.~\ref{sec:results}. In Sec.~\ref{sec:AdS} we
investigate the behaviour of the self-force in Schwarzschild-anti de
Sitter spacetime by formulating a simple approximation that can be
treated analytically, and showing that it reproduces our numerical
results when $r_0 \gg r_c$. We offer some concluding remarks in
Sec.~\ref{sec:conclusion}, and in the Appendix we explain why we
refrain from computing the self-force on a scalar charge in
Schwarzschild-de Sitter spacetimes.  

\section{Self-force computations}   
\label{sec:methods} 

\subsection{Spacetime} 

We work with the class of Schwarzschild-de Sitter spacetimes, with a
metric described by 
\begin{equation} 
ds^2 = -f\, dt^2 + f^{-1} dr^2 + r^2\, d\Omega^2, 
\label{eq:metric} 
\end{equation} 
with 
\begin{equation} 
f = 1 - \frac{2M}{r} - \frac{1}{3} \Lambda r^2 
\label{eq:f_def} 
\end{equation} 
and $d\Omega^2 = d\theta^2 + \sin^2\theta\, d\phi^2$. The spacetime
contains a black hole of mass $M$, and the metric is a solution to the
Einstein field equations with cosmological constant $\Lambda$. We
allow $\Lambda$ to be either positive or negative; when $\Lambda<0$
the solution is often referred to as the Schwarzschild-anti de Sitter
metric. 

When $\Lambda>0$ the spacetime contains both an event horizon at
$r=r_e$ and a cosmological horizon at $r=r_c$. It is useful to adopt
$r_e$ and $r_c$ as the primary parameters of the spacetime, and to
express $M$ and $\Lambda$ as functions of these parameters. This is
achieved by factorizing $f$ as $f = -k(r-r_e)(r-r_c)(r + r_e + r_c)/r$ and
comparing with Eq.~(\ref{eq:f_def}). We get 
\begin{equation} 
2M = \frac{r_e r_c(r_e+r_c)}{r_c^2+r_e r_c + r_e^2}, \qquad 
\Lambda = \frac{3}{r_c^2+r_e r_c + r_e^2}. 
\end{equation} 
When $r_e \ll r_c$ these relations reduce to $2M \simeq r_e$ and
$\Lambda \simeq 3/r_c^2$. 

When $\Lambda<0$ the spacetime still contains an event horizon at
$r=r_e$, but there is no longer a cosmological horizon. The spacetime
possesses anti de Sitter asymptotics, with a timelike conformal
boundary at $r=\infty$. In this case also it is useful to use $r_e$
and a cosmological length scale $r_c$ to parametrize the
spacetime. The definition of $r_c$ is somewhat arbitrary, and we fix
it with the factorization $f = -k(r-r_e)(r^2+r_e r + r_c^2)/r$. This
produces the alternative relations 
\begin{equation} 
2M = \frac{r_e r_c^2}{r_c^2-r_e^2}, \qquad 
\Lambda = -\frac{3}{r_c^2-r_e^2}
\end{equation} 
for the mass and cosmological constant. When $r_e \ll r_c$ we still
have $2M \simeq r_e$ and $\Lambda \simeq -3/r_c^2$. 

\subsection{Electromagnetic self-force} 

The equations that permit a computation of the self-force on a charge
at rest in a static, spherically-symmetric spacetimes were developed
in Sec.~X of Ref.~\cite{casals-poisson-vega:12}. We provide a brief
summary, specializing their results to the specific case of a
Schwarzschild-de Sitter spacetime. 

The electromagnetic field tensor $F_{\alpha\beta}$ is expressed as
$F_{\alpha\beta} = \nabla_\alpha A_\beta - \nabla_\beta A_\alpha$ in
terms of a potential $A_\alpha$, and in the Lorenz gauge
$\nabla_\alpha A^\alpha = 0$ the complete set of Maxwell equations
reduce to 
\begin{equation} 
\Box A^\alpha - R^{\alpha}_{\ \beta} A^\beta = -4\pi j^\alpha, 
\label{eq:wave_em} 
\end{equation} 
in which $\Box = g^{\alpha\beta} \nabla_\alpha \nabla_\beta$ is the
wave operator in curved spacetime, $R^\alpha_{\ \beta}$ is the Ricci
tensor, and $j^\alpha$ is the current density. For a point particle
moving on a world line described by the parametric equations 
$x^\alpha = z^\alpha(\tau)$, where $\tau$ is proper time, the current
density at the spacetime event $x$ is  
\begin{equation} 
j^\alpha(x) = e \int u^\alpha \delta\bigl(x,z(\tau)\bigr)\, d\tau, 
\end{equation} 
in which $e$ is the particle's electric charge, $u^\alpha =
dz^\alpha/d\tau$ is the velocity vector, and $\delta(x,x')$ is a
scalarized delta function. In our developments here, the particle is
placed on a static world line described by $r=r_0$. 

A computation of the self-force involves the component $A_t$ only, and
a development in spherical harmonics, 
\begin{equation} 
A_t(r,\theta,\phi) = \sum_{\ell m} R_{\ell m}(r)
Y_{\ell m}(\theta,\phi), 
\end{equation} 
turns Eq.~(\ref{eq:wave_em}) into an ordinary second-order
differential equation for the radial functions $R_{\ell m}$. By
placing the particle on the polar axis $\theta = 0$ we ensure that
modes with $m\neq 0$ vanish, and we obtain
\begin{equation} 
r^2 R_{\ell}'' + 2 r R_{\ell}' - \frac{\ell(\ell+1)}{f} R_{\ell} 
= 4\pi e \sqrt{\frac{2\ell+1}{4\pi}} \delta(r-r_0)  
\label{eq:diffeq_em} 
\end{equation} 
for the nonvanishing modes $R_\ell := R_{\ell 0}$; a prime indicates 
differentiation with respect to $r$. 

Near a horizon at $f=0$, Eq.~(\ref{eq:diffeq_em}) admits the
asymptotic expansion $R = R_{h} f (1 + R_1 f + R_2 f^2 + \cdots)$,
in which the coefficients $R_1$, $R_2$, and so on can be determined
from the differential equation; the overall normalization $R_{h}$
is arbitrary. This expression can be used to set the values of 
$R_{\ell}$ and $R_{\ell}'$ for a numerical integration of the
equation that starts at $r=r_e(1+\epsilon)$ with $\epsilon \ll 1$. The
strategy works also for an integration starting at $r=r_c(1-\epsilon)$,
when the spacetime contains a cosmological horizon. For anti de Sitter
asymptotics the differential equation produces the asymptotic
behaviours $e^{\pm s/r}$, where 
$s^2 := 3\ell(\ell+1)/|\Lambda|$. Both solutions are bounded 
in the limit $r \to \infty$, and to make the problem well-posed we
impose the Dirichlet boundary condition $A_t(\infty,\theta,\phi) 
= 0$ on the potential. (This can be thought of as the static limit of 
reflecting-wave boundary conditions at $r=\infty$.) The selected
asymptotic form is therefore $R \sim R_\infty (e^{s/r} - e^{-s/r})$,
and this can be used to set the values of $R_{\ell}$ and $R_{\ell}'$
for a numerical integration that starts at $r=r_\infty \gg r_c$; the
overall normalization $R_\infty$ is arbitrary. 

The integration of Eq.~(\ref{eq:diffeq_em}) proceeds outward from
$r=r_e(1+\epsilon)$ with an undetermined normalization $R_e$, and it
simultaneously proceeds inward from $r=r_c(1-\epsilon)$ or 
$r = r_\infty$ with another undetermined normalization $R_c$ or
$R_\infty$. The normalizations are determined by enforcing the
junction conditions 
\begin{equation} 
\bigl[ R_{\ell} \bigr] = 0, \qquad 
\bigl[ R'_{\ell} \bigr] = \frac{4\pi e}{r_0^2}
\sqrt{\frac{2\ell+1}{4\pi}} 
\end{equation} 
at $r=r_0$, where $[\psi] = \psi(r_0^+) - \psi(r_0^-)$ denotes the
jump of a quantity $\psi$ across $r=r_0$; in an obvious notation,
$r_0^\pm = r_0(1\pm\epsilon)$ with $\epsilon>0$, and the jump is
evaluated in the limit $\epsilon \to 0$. 

The equation for $\ell = 0$ admits an analytical solution. The general
solution to the homogeneous equation is $R_{0} = a + b/r$, where $a$
and $b$ is a constant. The constant solution has no impact on the
self-force, and enforcing the junctions conditions, we have that 
$R_{0} = -\sqrt{4\pi} e/r$ for $r > r_0$, and 
$R_{0} = -\sqrt{4\pi} e/r_0$ for $r < r_0$. 

The radial functions can next be involved in a mode-sum computation of
the electromagnetic self-force. The required expression can be
obtained from Eqs.~(10.10), (10.22), and (10.23) of
Ref.~\cite{casals-poisson-vega:12}. We have 
\begin{equation} 
\fl
F^r = e^2 f_0 \sum_{\ell = 0}^\infty \biggl[ 
\sqrt{\frac{2\ell+1}{4\pi}} \frac{R'_{\ell}(r_0 + \Delta)}{e f_0^{1/2}} 
- A \Bigl( \ell + \frac{1}{2} \Bigr) - B 
- \frac{D}{(\ell-\frac{1}{2})(\ell+\frac{3}{2})} 
- \cdots \biggr], 
\label{eq:sf_em} 
\end{equation} 
in which $f_0 = f(r_0)$ and $A$, $B$, and $D$ are regularization
parameters that permit the mode-by-mode subtraction of the singular
part of the electromagnetic field (the method originates in the work
of Barack and Ori; see Refs.~\cite{barack-etal:02, barack-ori:02,
  barack-ori:03a}). Explicit expressions are given in Eq.~(10.24) of
Ref.~\cite{casals-poisson-vega:12}, and for the specific case of
Schwarzschild-de Sitter spacetime the regularization parameters are
given by   
\begin{eqnarray} 
\fl
A &=& \frac{1}{r_0^2 f_0^{1/2}} \mbox{sign}(\Delta), \\ 
\fl
B &=& -\frac{1}{2 r_0^3 f_0} (3M - r_0), \\ 
\fl
D &=& \frac{1}{48r_0^5 f_0^2} \bigl( 45M^3 - 27M^2 r_0 
+ 24M^2 \Lambda r_0^3 - 9 M \Lambda r_0^4 
+ 3\Lambda r_0^5 - 4M\Lambda^2 r_0^6 \bigr). 
\end{eqnarray} 
In Eq.~(\ref{eq:sf_em}) the derivative of the radial function is
evaluated at $r = r_0 + \Delta$ in the limit $\Delta \to 0$; the limit
can be taken from either direction, and the sign of $A$ reflects this
choice. 

\subsection{Numerical implementation} 

The differential equation (\ref{eq:diffeq_em}) was integrated
numerically using the routine {\tt odeint} of the SciPy (Scientific
Python) library. For Schwarzschild-de Sitter spacetime the equation is
integrated forward from $r = r_e(1+\epsilon)$ and backward from  
$r = r_c(1-\epsilon)$, using starting values obtained from an 
asymptotic analysis. For Schwarzschild-anti de Sitter spacetime the
backward integration proceeds instead from $r = r_\infty \gg r_c$,
with appropriate starting values. In both cases the solutions are
matched at $r=r_0$ by enforcing the junction conditions. 

The radial functions $R'_{\ell}(r_0)$ are then inserted within
Eq.~(\ref{eq:sf_em}) to evaluate the electromagnetic self-force. The
regularization procedure, which involves subtracting the $A$, $B$, and
$D$ terms from the radial function, is a powerful diagnostic of
numerical accuracy. Examining Eq.~(\ref{eq:sf_em}), we observe that a
plot of the bracketed quantity as a function of $\ell$ must reveal a
$\ell^{-4}$ falloff when $\ell$ is large; failure to obtain this
behaviour signals either coding errors or issues of numerical
stability. All results presented below have been validated by a
careful monitoring of such multipole plots.     

The mode sum of Eqs.~(\ref{eq:sf_em}) is typically truncated at 
$\ell = \ell_{\rm max} = 80$. Additional accuracy can be obtained by
estimating the remaining portion of the sum
\cite{diaz-rivera-etal:04}. If we denote each term in the mode sum by
$F^r_\ell$, then the self-force is calculated as $F^r = \sf{data} 
+ \sf{tail}$, with $\sf{data}$ given by the partial mode sum up to 
$\ell = \ell_{\rm max}$, and $\sf{tail}$ by the remaining sum from
$\ell_{\rm max} + 1$ to infinity. There is no numerical data to
calculate the tail, but when $\ell_{\rm max}$ is sufficiently large,
$F^r_\ell$ can be approximated by \cite{diaz-rivera-etal:04} 
\[
\frac{E}{(\ell-\frac{3}{2})(\ell-\frac{1}{2})(\ell+\frac{3}{2}) 
(\ell+\frac{5}{2})}.  
\]
Summing over this estimate yields 
\begin{equation} 
{\sf tail} \simeq \frac{16E}{3} 
\frac{\ell_{\rm max}+1}{(2 \ell_{\rm max}-1)(2 \ell_{\rm max}+1)
(2 \ell_{\rm max}+3)(\ell_{\rm max}+5)}. 
\end{equation} 
The coefficient $E$ is unknown, but it can be estimated from the last
data point at $\ell = \ell_{\rm max}$,  
\begin{equation}  
E \simeq {\textstyle (\ell_{\rm max}-\frac{3}{2})
(\ell_{\rm max}-\frac{1}{2})(\ell_{\rm max}+\frac{3}{2})
(\ell_{\rm max}+\frac{5}{2})} F^r_{\ell_{\rm max}}. 
\end{equation} 
Making the substitution returns the estimate 
\begin{equation} 
{\sf tail} \simeq \frac{(2 \ell_{\rm max}-3)(\ell_{\rm max}+1)}
{3(2 \ell_{\rm max}+1)} F^r_{\ell_{\rm max}} 
\end{equation} 
for the tail. Adding this to $\sf{data}$ improves the final result for
the self-force, and provides a gross overestimate of the numerical
accuracy. With $\ell_{\rm max} = 80$ we find that the error bars on
$F^r$ would not be visible on the plots displayed in
Sec.~\ref{sec:results}. A more accurate estimation of ${\sf tail}$ 
\cite{diaz-rivera-etal:04} uses additional points near 
$\ell = \ell_{\rm max}$ to fit for additional parameters $F$, $G$,
$H$, which are then included with appropriate polynomials in 
${\sf tail}$; we do not pursue such refinements here.   

\section{Self-force results} 
\label{sec:results} 

\subsection{Schwarzschild-de Sitter spacetime} 
\label{sec:results_SdS} 

We have performed a range of self-force computations for a static
particle in Schwarzschild-de Sitter spacetime. The computations have
three input parameters: the event-horizon radius $r_e$, the particle's
position $r_0$, and the cosmological radius $r_c$; these are
restricted by $r_e < r_0 < r_c$. It is convenient to rescale all
lengths by $r_e$, which can be thought of as the fundamental scale of
the spacetime. Because $r_e \simeq 2M$ when $r_e \ll r_c$, we find it
helpful to set $r_e = 2$ in our computations, loosely thinking of the
mass of the black hole as setting the unit of length.  The particle's
charge $e$ is not an essential input parameter, because the self-force
necessarily scales as $e^2$; for our computations we set $e=1$.  

\begin{figure}
\includegraphics[width=5in]{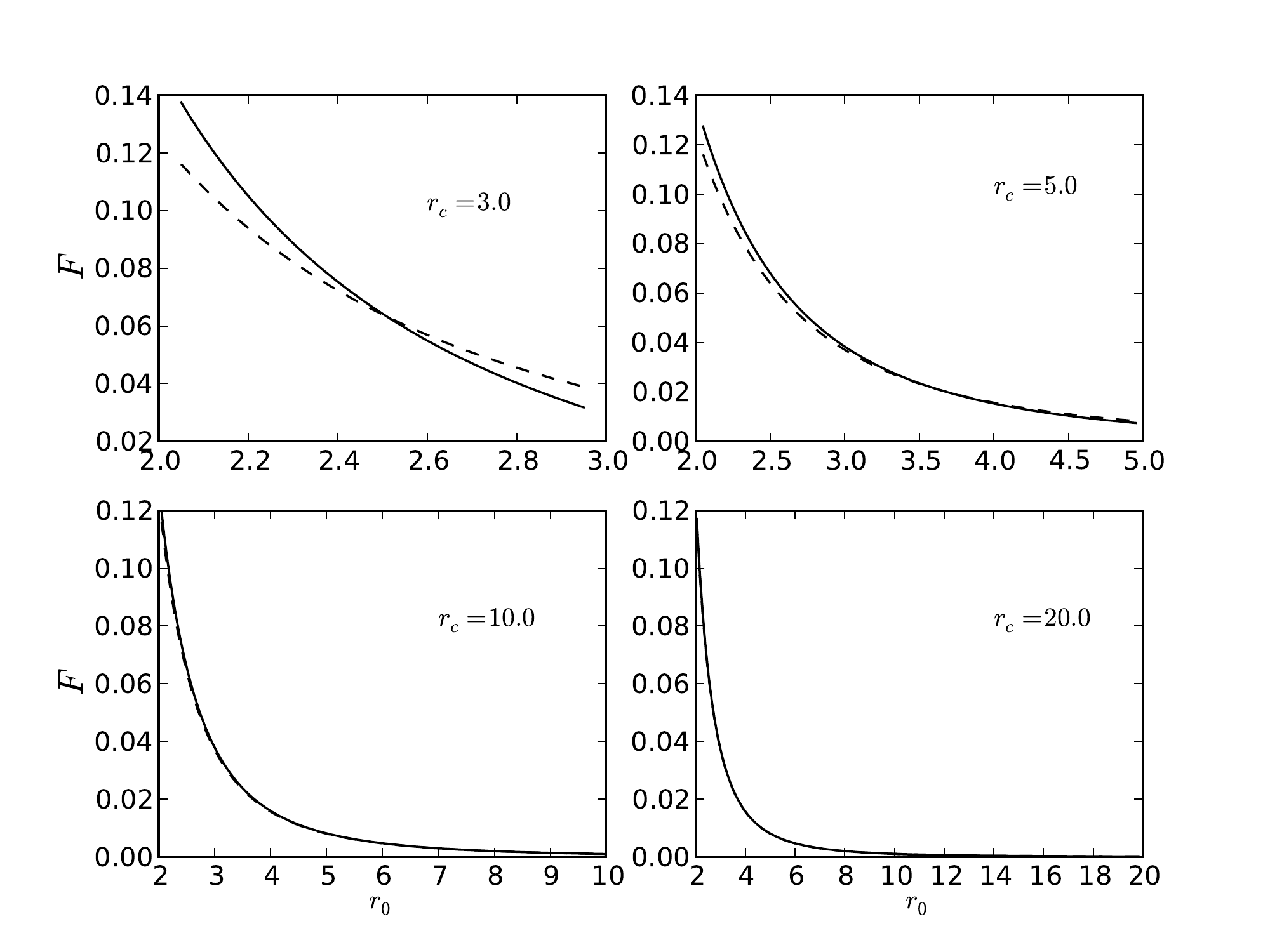}
\caption{Electromagnetic self-force $F = F^r/\sqrt{f_0}$ acting on a
  static charge in Schwarzschild-de Sitter spacetime. In all cases
  $r_e = 2$ in arbitrary units, and computations are shown as a solid
  curve for selected values of $r_c$. The dashed curve represents the
  Smith-Will force for a black hole of mass $M=1$. The solid and
  dashed curves are barely distinguishable when $r_c = 10$ and $r_c =
  20$.} 
\label{fig:1}
\end{figure}

A sample of our results are presented in Fig.~\ref{fig:1}; for each
selected value of $r_c$ we display $F := f_0^{-1/2} F^r$ as a function
of $r_0$. The division by $\sqrt{f_0}$, where $f_0 := f(r_0)$,
translates $F^r$ into a coordinate-invariant quantity. For example,
$F$ is the self-force as measured by a static observer at $r=r_0$,
making use of an orthonormal frame there. Alternatively, $F$ is the
square root of the spacetime invariant 
$g_{\alpha\beta} F^\alpha F^\beta$, with a choice of sign
inherited from $F^r$. An immediate observation from Fig.~\ref{fig:1}
is that $F > 0$ in all cases; the electromagnetic self-force always
represents a repulsion from the black hole. A second observation is
that the self-force is monotonically decreasing as a function of
$r_0$.  

To reflect on these results it is helpful to begin with cases for
which $r_c \gg r_e$. In such situations we can expect that an electric
charge at $r_0 \ll r_c$ will feel a self-force that is mostly
sensitive to the black-hole aspects of the spacetime, and largely
insensitive to the de Sitter asymptotic conditions. We would therefore
expect the self-force to be well approximated by the Smith-Will
expression \cite{smith-will:80} 
\begin{equation} 
F = \frac{e^2 M}{r_0^3},   
\label{eq:SW} 
\end{equation} 
in which $M$ is well approximated by $r_e/2$. On the other hand, an
electric charge at $r_0 \gg r_e$ should feel a self-force
that is mostly sensitive to the de Sitter aspects of the
spacetime. Because de Sitter spacetime is conformally flat, and
because Maxwell's equations are conformally invariant, the self-force
should vanish in this limit. Combining these observations, we would
expect that the self-force should display a change of behaviour as
$r_0$ increases from $r_e$ toward $r_c$; the Smith-Will behaviour
should gradually give way to a self-force that approaches zero
faster than $r_0^{-3}$. 

These expectations are borne out by the numerical results. We see
generally speaking that when $r_e \ll r_c$, the self-force is well
approximated by the Smith-Will expression when $r_0$ is small, and
that the force becomes smaller than this when $r_0$ becomes comparable
to $r_c$. The difference between $F$ and the Smith-Will force is very
slight when $r_c \gg r_e$, but it increases as $r_c$ is decreased
toward $r_e$; in this regime there is no longer a clean separation of
scales, and the self-force is sensitive to all aspects of the
spacetime. As $r_c$ becomes comparable to $r_e$ we see that the
Smith-Will expression underestimates the self-force when $r_0$ is
close to $r_e$, and overestimates it when $r_0$ is close to $r_c$.
The discrepancy, however, is never worse than approximately 15\%,
which allows to conclude that the Smith-Will expression always
provides an adequate approximation to the self-force.  

\subsection{Schwarzschild-anti de Sitter spacetime} 
\label{sec:results_SadS} 

\begin{figure}
\includegraphics[width=5in]{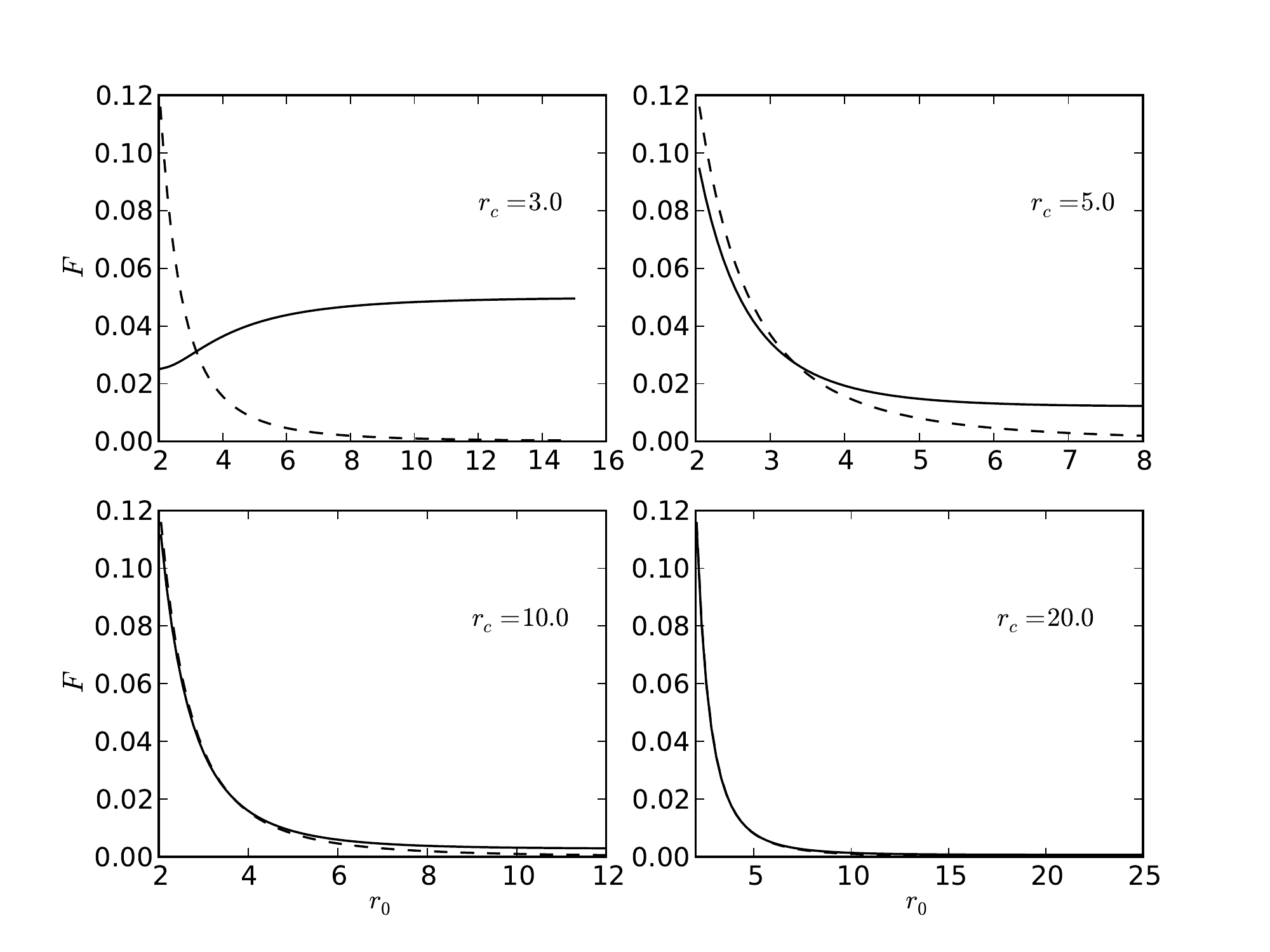}
\caption{Electromagnetic self-force $F = F^r/\sqrt{f_0}$ acting on a
  static charge in Schwarzschild-anti de Sitter spacetime. In all cases
  $r_e = 2$ in arbitrary units, and computations are shown as a solid
  curve for selected values of $r_c \geq 3$. The dashed curve
  represents the Smith-Will force for a black hole of mass $M=1$. The
  solid and dashed curves are barely distinguishable when $r_c = 20$.} 
\label{fig:2}
\end{figure}
 
We have also performed a range of self-force computations for a static 
charge in Schwarzschild-anti de Sitter spacetime. Here also the
computations have three input parameters: the event-horizon radius
$r_e$, the particle's position $r_0$, and the cosmological length
scale $r_c$; in this case it is possible for $r_0$ to exceed $r_c$. 
Once more we set $r_e = 2$ in our computations.

\begin{figure}
\includegraphics[width=5in]{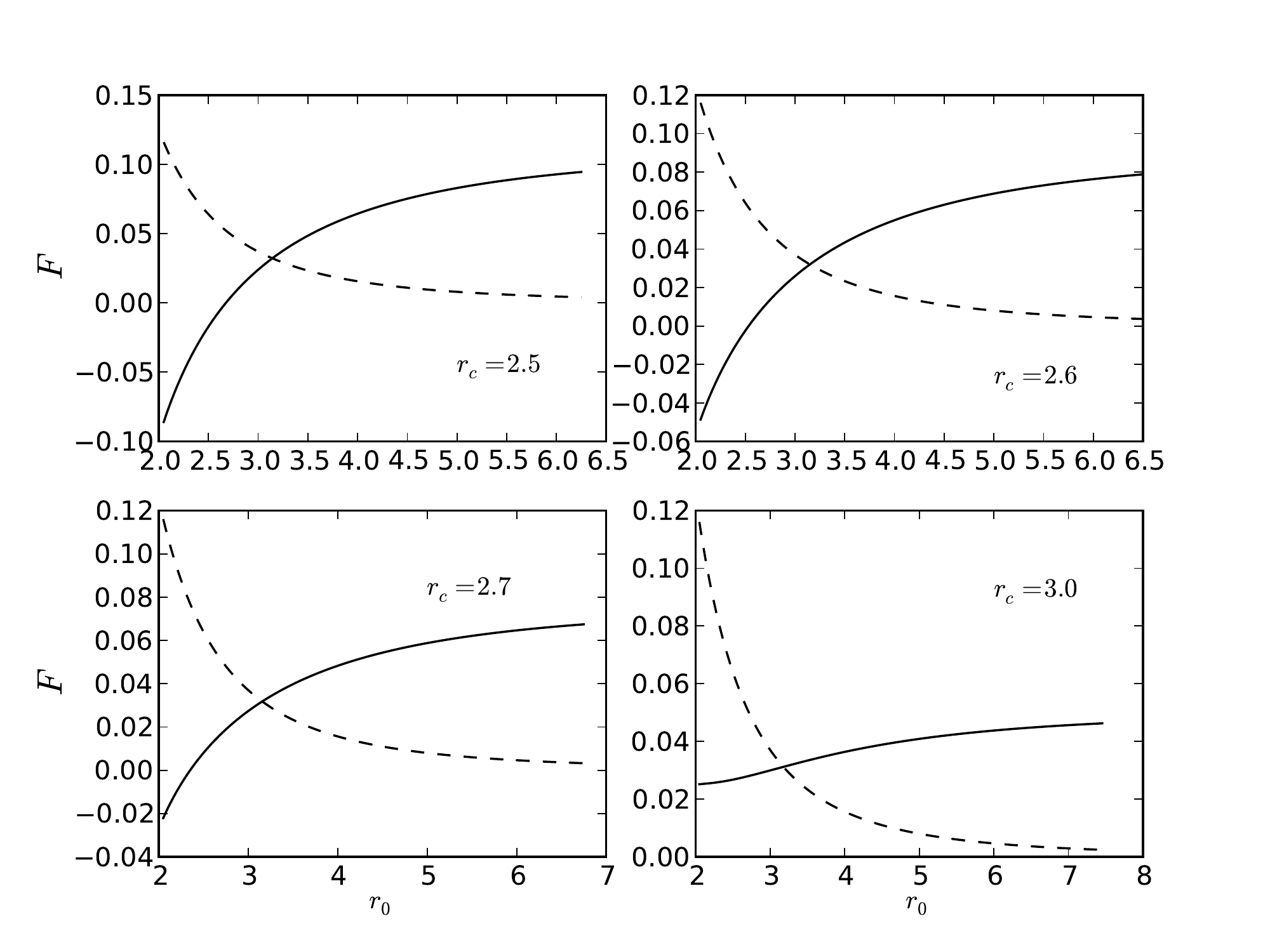}
\caption{Electromagnetic self-force acting on a static charge in 
  Schwarzschild-anti de Sitter spacetime. In all cases $r_e = 2$ in
  arbitrary units, and computations are shown for $r_c \leq 3$.} 
\label{fig:3}
\end{figure}
 
A sample of our results are presented in Figs.~\ref{fig:2} and
\ref{fig:3}. The figures reveal the following features. First, $F > 0$
whenever $r_c > 3$, so that in this regime the self-force represents a
repulsion from the black hole. Second, we notice that $F < 0$ when
$r_0 < r_c < 2.9$, so that in this regime the self-force actually
represents an attraction toward the black hole. Third, the self-force
is well approximated by the Smith-Will expression when 
$r_c \gg r_e$, but only when $r_0$ is smaller than $r_c$; when $r_0 >
r_c$ we see the self-force approaching a constant asymptotic value
$F_\infty$, while the Smith-Will expression continues to decay as
$r_0^{-3}$. Fourth, when $r_c$ is comparable to $r_e$, the Smith-Will
expression significantly overestimates the self-force when $r_0 <
r_c$, and it significantly underestimates it when $r_0 > r_c$. And
fifth, when $r_c > 4.9$ the self-force is monotonically 
decreasing toward $F_\infty$ as $r_0$ increases toward infinity, while
it is monotonically increasing toward $F_\infty$ when $r_c < 3.0$;
when $3.0 < r_c < 4.9$ the self-force is no longer monotonic, reaching
a minimum value at some $r_0$ before increasing again (this behaviour
is not shown in the figures). 

The existence of a regime $r_0 < r_c < 2.9$ giving rise to an attractive
self-force is a surprising feature of Schwarzschild-anti de Sitter
spacetime; there is no intuitive explanation for this outcome. Another
surprising feature is the approach to a constant asymptotic value
$F_\infty$ when $r_0 \gg r_c$. Our numerical results indicate 
that when $r_e \ll r_c$, $F_\infty$ is extremely well approximated by
the expression  
\begin{equation} 
F_\infty = \frac{e^2}{4 r_c^2}, 
\label{eq:Finf} 
\end{equation} 
so that it decreases with increasing $r_c$ --- this behaviour can be
gleaned from Fig.~\ref{fig:2}. As we shall see in Sec.~\ref{sec:AdS},
it is actually easy, after the fact, to explain this result. Another
regime that comforts intuition is the one with $r_e \ll r_c$ and 
$r_0 \ll r_c$, in which the self-force is dominated by the black-hole
aspects of the spacetime, and well approximated by the Smith-Will
expression.   

\section{Electromagnetic self-force in anti de Sitter spacetime} 
\label{sec:AdS} 

In this section we shed some light on the asymptotic limit of
Eq.~(\ref{eq:Finf}) by considering a static charge at a radius 
$r_0 \gg r_c$ in Schwarzschild-anti de Sitter spacetime. In this
portion of the spacetime the metric is well approximated by 
\begin{equation} 
ds^2 \simeq \frac{r^2}{a^2} \biggl( -dt^2 + \frac{a^4}{r^2}\, dr^2 
+ a^2\, d\Omega^2 \biggr), 
\end{equation} 
where $a^2 := 3/|\Lambda|$; when $r_e \ll r_c$ we have that 
$a \simeq r_c$. The coordinate transformation $\rho = a(1-a/r)$ brings
the metric to the form 
\begin{equation} 
ds^2 \simeq \frac{r^2}{a^2} \Bigl(-dt^2 + d\rho^2 
+ a^2\, d\Omega^2 \Bigr), 
\end{equation} 
in which $r$ is to be expressed as a function of $\rho$. The spacetime
is approximately conformal to a spacetime with metric 
\begin{equation} 
d\tilde{s}^2 = -dt^2 + d\rho^2 + a^2\, d\Omega^2, 
\end{equation} 
and the self-force on the static charge can be calculated by working
in the conformal spacetime. 

The conformal boundary is situated at $\rho = a$, and the conformal
spacetime is necessarily truncated at this limiting value. The
Dirichlet condition $A_t(r=\infty,\theta,\phi) = 0$ imposed 
on the potential implies that the boundary acts as a grounded
conductor. When $r_0 \gg a$ we have that $\rho_0 = a(1-a/r_0)$
is very close to $a$, so that the charge is in the immediate
vicinity of the conductor. The electric field produced by the charge
is therefore strongly influenced by the conducting surface, and we
expect that it can be computed by exploiting the method of images. We
also expect the self-force to arise as a consequence of the interaction
between the charge and its image.    

To carry out the calculations we work in a small coordinate patch
around the charge. The patch is centered at $(\rho_0, \theta_0,
\phi_0)$, and because the $\rho$-dependence of the electric field is
the central aspect of the problem, we allow ourselves to approximate
the angular part of the line element as 
\begin{equation} 
a^2\, d\Omega^2 \simeq a^2\, d(\theta-\theta_0)^2 
+ a^2\sin^2\theta_0\, d(\phi-\phi_0)^2 
= dx^2 + dy^2, 
\end{equation} 
where $x = a(\theta-\theta_0)$ and $y =
a\sin\theta_0(\phi-\phi_0)$ are locally Cartesian coordinates. With
this simplification the conformal metric becomes 
\begin{equation}
d\tilde{s}^2 = -dt^2 + \rho^2 + dx^2 + dy^2, 
\end{equation} 
and the conformal boundary at $\rho = a$ becomes flat. 

This situation is now elementary. We have a charge $e$ at $\rho =
\rho_0$, a flat conductor at $\rho = a$, and an image charge $-e$ at
a position $\rho = a + (a-\rho_0)$ beyond the conductor. The
distance between the charges is $d = 2(a-\rho_0) = 2a^2/r_0$, and
the electric field supplied by the image charge is equal to 
$E = e/d^2 = e r_0^2/(4 a^4)$ when evaluated at the physical charge; 
the field points in the positive $\rho$ direction, toward the image
charge. The electromagnetic field tensor is $\tilde{F}_{t\rho} = -E$
in the conformal spacetime.  

Because Maxwell's equations are conformally invariant, we have that   
$F_{\alpha\beta} = \tilde{F}_{\alpha\beta}$, and our previous 
results imply that $E = -F_{t\rho}$ is also the electric field in the
physical spacetime. Transforming back to the original radial
coordinate, we have that 
\begin{equation} 
F_{tr} = \frac{d\rho}{dr} F_{t\rho} = -\frac{e}{4a^2}.  
\end{equation} 
The force $F^\alpha = e F^\alpha_{\ \beta} u^\beta$ exerted by
this electric field evaluates to $F^r = e^2 f_0^{1/2}/(4 a^2)$, 
or 
\begin{equation} 
F = \frac{e^2}{4 a^2} 
\end{equation} 
after division by $\sqrt{f_0}$ to convert $F^r$ into the invariant
$F$. When $r_e \ll r_c$ this reduces to $F = e^2/(4 r_c^2)$, and this
is precisely what was observed in Sec.~\ref{sec:results_SadS}. We see
that the constant asymptotic behaviour is explained very convincingly
in terms of an interaction between the charge and the conducting
surface that must be placed at the conformal boundary to enforce the
Dirichlet boundary conditions. While the relation of
Eq.~(\ref{eq:Finf}) was first obtained by means of a fit to our
numerical results, we see that the value of $F_\infty$ is revealed by
a very simple analytical calculation. 

\section{Concluding remarks} 
\label{sec:conclusion} 

We have computed the self-force acting on an electric charge at rest
in Schwarzschild-de Sitter spacetimes, allowing the cosmological
constant to be either positive or negative. Our results reveal some
intuitive features, but they also display aspects that defy
intuition. 

In the case of a positive cosmological constant, we have seen that the
self-force is always positive, representing a repulsion from the black
hole, and monotonically decreasing with increasing $r_0$. When 
$r_e \ll r_c$, so that the black-hole and cosmological scales are well 
separated, the self-force is dominated by the black-hole aspects of
the spacetime when $r_0 \ll r_c$, and it is  well approximated by the
Smith-Will expression of Eq.~(\ref{eq:SW}); it falls off faster than
the Smith-Will expression when $r_0 \gg r_e$, where the de Sitter
aspects of the spacetime are dominant. These results are intuitive, to
the extent that they can be expected on the basis of the Smith-Will
force (which is itself unintuitive) and the fact that the self-force
necessarily vanishes in pure de Sitter spacetime. When $r_e$ is
comparable to $r_c$, the mixing of scales makes this intuition
unreliable. The self-force, however, continues to behave in a
qualitatively similar manner, and we note that the Smith-Will
expression continues to provide an adequate approximation; the
discrepancy never exceeds 15\% in the full range $r_e < r_0 < r_c$.  

The spectrum of results is richer in the case of a negative
cosmological constant. Here the self-force is not always positive --- 
it goes negative when $r_0 < r_c < 3$, in units in which $r_e = 2$ ---
and it is monotonically decreasing for $r_c > 4.9$ only --- it is  
monotonically increasing for $r_c < 3.0$. These features are not 
intuitive. One aspect that could be anticipated, however, is the  
asymptotic approach to a constant value $F_\infty$ when $r_0 \gg r_c$;
as we saw in Sec.~\ref{sec:AdS}, this is the result of an interaction
between the charge and the conformal boundary at $r = \infty$, which
acts as a grounded conductor.   

It is comforting that some aspects of the self-force admit intuitive
explanations, but it is perplexing that so many do not. This situation
continues to motivate further study. 

\appendix 

\section{Scalar self-force} 

The self-force on a scalar charge $q$ in curved spacetime is often
considered as a toy model for the electromagnetic self-force, and in
most situations the computations of the scalar self-force are
considerably simpler. For a charge at rest in a static spacetime the
degree of complexity is exactly the same, because as we have seen, the
computation of the electromagnetic self-force involves a single
component of the vector potential $A_\alpha$. So with the
infrastructure put in place to calculate the electromagnetic
self-force, it is usually a simple matter to make the changes required
to compute a scalar self-force.

Our intention was indeed to perform scalar self-force computations,
until we came to realize that a static scalar charge in
Schwarzschild-de Sitter spacetime produces a scalar field that is
necessarily singular at one of the horizons. The conclusion extends to
pure de Sitter spacetime, where the field either diverges at $r=0$ or
at $r=r_c$. The conclusion, however, does not extend to
Schwarzschild-anti de Sitter spacetime, to pure anti de Sitter
spacetime, or to a scalar field that is nonminimally coupled to the
spacetime curvature. Because of these issues, we have abandoned our
goal to compute scalar self-forces in Schwarzschild-de Sitter
spacetimes.  

The problem is easily identified with an examination of the $\ell = 0$ 
mode of the scalar equation. The scalar field $\Phi$ is taken to
satisfy the wave equation  
\begin{equation} 
\Box \Phi = -4\pi \mu, 
\label{eq:wave_scalar} 
\end{equation} 
in which $\mu$ is the scalar charge density. For a point particle this
is given by 
\begin{equation} 
\mu(x) = q \int \delta\bigl(x, z(\tau)\bigr)\, d\tau, 
\end{equation} 
where $q$ is the particle's scalar charge. The particle is placed on a
static world line $z(\tau)$ described by $r=r_0$. A decomposition in
spherical harmonics,  
\begin{equation} 
\Phi(r,\theta,\phi) = \sum_{\ell m} R_{\ell m}(r)
Y_{\ell m}(\theta,\phi), 
\end{equation} 
gives rise to the ordinary second-order differential equation  
\begin{equation} 
\fl
r^2 R_{\ell}'' + \biggl( 2 + \frac{rf'}{f} \biggr) r R_{\ell}' 
- \frac{\ell(\ell+1)}{f} R_{\ell} 
= -\frac{4\pi q}{f_0^{1/2}} \sqrt{\frac{2\ell+1}{4\pi}} \delta(r-r_0)   
\label{eq:diffeq_scalar} 
\end{equation} 
for the nonvanishing modes $R_\ell := R_{\ell 0}$; a prime indicates 
differentiation with respect to $r$. The radial functions are required
to satisfy the junction conditions 
\begin{equation} 
\bigl[ R_{\ell} \bigr] = 0, \qquad 
\bigl[ R'_{\ell} \bigr] = -\frac{4\pi e}{r_0^2 f_0^{1/2}}
\sqrt{\frac{2\ell+1}{4\pi}} 
\end{equation} 
at $r=r_0$.

The equation for $\ell = 0$ admits an analytical solution. The
linearly independent solutions are $dR_{0}/dr = 0$ and 
$dR_{0}/dr = c/(r^2 f)$, and the constant $c$ is determined by the
junction conditions. The source of our troubles is the factor of $f$
in the denominator. Because $f = 0$ at either horizon, a solution that
is regular at $r=r_e$ is necessarily singular at $r=r_c$, and vice
versa. In pure de Sitter spacetime, the field can be made regular at
$r=r_c$ by adopting the zero solution for $r > r_0$, but the internal
solution will then diverge as $r^{-2}$ near $r=0$. These problems go
away when there is no cosmological horizon; with anti de Sitter
asymptotics the external solution can be set equal to $c/(r^2 f)$,
which decays as $r^{-4}$ when $r \to \infty$. 

Because the singularity in the $\ell = 0$ mode of the scalar field 
cannot be cured by the other modes, we conclude that a static scalar 
charge produces a field that necessarily diverges at one of the
horizons in Schwarzschild-de Sitter spacetime. As we have seen, the
pathology extends to pure de Sitter spacetime, but it is avoided in
Schwarzschild-anti de Sitter and pure anti-de Sitter
spacetimes. Changing the scalar-field equation by introducing a
nonminimal coupling to the curvature also offers a way out. 

\ack

This work was supported by the Natural Sciences and Engineering
Research Council of Canada. 

\section*{References}
\bibliography{../bib/master} 
\end{document}